\documentclass[12pt,preprint]{aastex}

\newcommand{\beq}{\begin{equation}}
\newcommand{\eeq}{\end{equation}}
\newcommand{\Mdot}{\dot{M}~}

\newcommand{\kms}{\mbox{ km s$^{-1}$}~}
\newcommand{\gcms}{\mbox{ g cm s$^{-1}$}~}
\newcommand{\gcm}{\mbox{ g cm$^{-3}$}~}

\newcommand{\Mo}{\mbox{M$_{\odot}$}~}
\newcommand{\Moy}{\mbox{M$_{\odot}$ yr$^{-1}$}~}

\begin{document}
 
\def\etal{{\it et~al.\ }}
\def\eg{{\it e.~g.\ }}
\def\ie{{\it i.~e.,\ }}
 
\title{Magnetic-Driven Winds from Post-AGB Stars: \\
Solutions for High Speed Winds and Extreme Collimation} 

\author{Guillermo Garc\'{\i}a-Segura\altaffilmark{1}, 
Jos\'e Alberto L\'opez\altaffilmark{1}
and Jos\'e Franco\altaffilmark{2}}
\affil{Instituto de Astronom\'{\i}a-UNAM}
\altaffiltext{1}{Instituto de Astronom\'{\i}a-UNAM, Apdo Postal 877,
Ensenada 22800, Baja California, Mexico; Email address: ggs@astrosen.unam.mx;
Email address: jal@astrosen.unam.mx}
\altaffiltext{2}{Instituto de Astronom\'{\i}a-UNAM, Apdo Postal 70-264, 
M\'exico 04510, M\'exico D. F., Mexico; Email address: pepe@astroscu.unam.mx}

\begin{abstract}
\noindent

This paper explores the effects of post-AGB winds driven solely by magnetic
pressure from the stellar surface. It is found that winds can reach high speeds
under this assumption, and lead to the formation of highly collimated 
proto-planetary nebulae. Bipolar knotty jets with periodic features and constant
velocity are well reproduced by the models. Several wind models with terminal
velocities from a few tens of $\kms$ up to $10^3$ $\kms$ are calculated,
yielding outflows with linear momenta in the range $10^{36}-10^{40} \gcms$, and
kinetic energies in the range $10^{42}-10^{47} $ erg. These results are in
accord with recent observations of proto-planetary nebulae that have pointed out
serious energy and momentum deficits if radiation pressure is considered as the
only driver for these outflows. Our models strengthen the notion that the large
mass-loss rates of post-AGB stars, together with the short transition times from
the late AGB to the planetary nebula stage, could be directly linked with the
generation of strong magnetic fields during this transition stage.
\end{abstract}

\keywords{Hydrodynamics---Stars: AGB, Post-AGB---ISM: Planetary Nebulae, 
Proto-Planetary Nebulae---ISM: Individual (OH 231.8+4.2, Mz-3, M 2-9, Hen 2-90, 
He 3-401)}

\vfill\eject
\section{Introduction} 

Winds from asymptotic giant branch (AGB) stars are thought to be driven by
radiation pressure on dust grains (see review by  Habing 1996), although an
alternative physical mechanism has been proposed by Pascoli (1997) based in
magnetic pressure that is transported out from the stellar interior to the
stellar surface. On the other hand, it is widely accepted that planetary
nebulae (PNs) are powered by line-driven winds emerging from their central
stars, and they are formed from a two-wind dynamic interaction (i.e., Kwok,
Purton \& Fitzgerald 1978). Evidences for this scenario are the large number 
of P-cygni line profiles detected in their central objects (Perinotto 1983).

Post-AGB stars with their associated Proto-planetary nebulae (PPNs) are 
short-lived transition objects between AGB stars and white dwarfs. Their 
energy source, although unclear, has been usually assumed to be radiation
pressure. However, recent observations of PPNs (Alcolea et al. 2001; Bujarrabal
et al. 2001 and references therein) have revealed that the linear momenta and
kinetic energies associated with these objects are in excess to what can be
provided by radiation pressure alone, in some cases by up to three orders of
magnitude. These large amounts of momentum and energy, as discussed in detail 
by Bujarrabal et al.(2001), cannot be accounted for by either radiation pressure
on dust grains, line-driven winds or continuum-driven winds.

The results discussed by Pascoli (1997), based on surface magnetic pressure as
the main driver of the large mass-loss rates in AGB stars, are an attractive
alternative to generate the required mechanical power in the winds of Post-AGB
stars, provided that the generation of magnetic fields can be efficient in 
post-AGB stars, as suggested by Blackman et al. (2001). Important indications
that this may actually be happening in some objects are found in the OH maser
radio observations of the PPN K 3-35 by Miranda et al. (2001), revealing the
existence of milligauss strength fields in a magnetized torus around the PPN.
Also, radio observations of PPNs CRL 2688 and NGC 7027 by Greaves (2002) show
evidence for the presence of toroidal magnetic fields. A closely related result
is that some nebulae, like Menzel 3 (Santander-Garc\'{\i}a et al. 2004), 
show an increasing degree of collimation with time, indicating that the 
internal $B$-field may get stronger as the evolution proceeds. 
These observational data indicate that magnetic-driven outflows, as those 
discussed by Pascoli, could be a
plausible mechanism to power the winds from Post-AGB stars. 
In this case, in order to make a quick transition from the late-AGB to the 
post-AGB stages, the stellar wind should sense a steep increase in the 
magnetic field strength that would allow the star to blow away a good 
fraction of its remaining envelope. 

There is not yet a clear model of how a single star can achieve this. 
One plausible scheme is that the rotation rate and the field strength at the
stellar core increase during the formation of the white dwarf. Thus, the inner
magnetic field becomes stronger as the core contracts and becomes exposed at
the stellar surface when the envelope is peeled-off during the PN formation.
Thus, a strong and dominant toroidal component develops at the interface 
between the core and the envelope, where some dynamo
action is expected and which may be responsible for launching a magnetic-driven
wind. Actually, Blackman et al. (2001) and Matt et al. (2004) have proposed that
the post-AGB wind is produced by magneto-centrifugal processes when dynamo
activity increases the internal field (see Blackman 2004), and the AGB star
sheds its outer layers, exposing the rotating and magnetized core. 
Obviously, more detailed stellar interior studies with rotation and $B$-fields 
are needed to
understand the details of this issue. In addition, some authors have suggested
that this may also occur in binary systems and, for instance, Soker (1997)
proposed that the Post-AGB stellar core can be spun-up by a secondary,
increasing the shear between the core and the envelope. Similarly, the spiral-in
process of a secondary star, or a giant planet, may also be able to produce a
large shear in the stellar envelope, and raise the magnetic field strength by
dynamo activity (Tout \& Reg\"os 2003). These cases link the large mass-loss
rates in the post-AGB stages with a common envelope phase. Unfortunately, there
are very few detailed studies of post-common envelope systems (e.g. Exter,
Pollaco \& Bell 2003). 
Here we do not select any particular scenario for creating a strong
toroidal stellar field, and focus our study only in what occurs after a strong
magnetized wind is created.

In previous papers \citep{RF96,GS97,GSetal99,GSL00}, we have assumed magnetized,
line-driven winds in the range $10^2-10^3 \kms $ with radial magnetic fields at the stellar surface. As in the case of the Solar wind, the magnetic fields are
primarily poloidal and parallel to the flow at small distances from the star,
stellar rotation creates a toroidal component and the hoop stress only plays an important role at large distances from the star. These solutions seem
appropriate to model PN with hot central stars, where line-driven winds should
be operative. In this paper we simply consider that the surface magnetic fields
are increased during the transition towards the Post-AGB stage and investigate
the effects of mass-loss when this is dominated by magnetic pressure, as 
suggested by Pascoli (1997). Here the wind is computed from the stellar surface,
and it is solved from sonic velocity up to the terminal velocity. 
For simplicity, the toroidal magnetic field at the stellar surface is the 
main and only driver
of the wind. This is an idealized, simple case, but it allows us to study the
effects of a magnetic-driven wind. \S 2 describes the numerical procedure. \S 3
presents and discusses the numerical results, and \S 4 presents the conclusions. 

\section{Numerical Models} 

A novel aspect in this paper is that the
stellar wind is not imposed at the inner boundary, instead it is computed with 
a simple scheme. We set an initial cold (100 K), isothermal atmosphere with a
power-law density stratification $\rho \sim r^{-2}$, corresponding to a
spherically symmetric wind with $\Mdot = 1 \times 10^{-6} \Moy$ and $v_{\infty}
= 10 \kms$. These are just reference values in order to compute the density.
The stellar gravitational field is always included as an external force which
corresponds to a point mass of $1 \Mo$.
We set outward sonic velocity to the wind and to the stellar surface
(i. e., the initial conditions are
almost static and, without any additional force, the whole atmosphere would
collapse in a free fall time scale). Once the grid is filled up with these
conditions, a toroidal magnetic field is introduced at the stellar surface with
a latitude dependence of the form
\beq
B_{\phi}(\theta) = B_{\rm s} \,\,\, {\rm sin} \, \theta ,
\eeq
where $ B_{\rm s}$ is the field at the equator ($\theta = \pi /2$).
In contrast with
our previous papers, where the surface magnetic field corresponds to the radial
component of the field, here we follow Pascoli (1997) and $B_{\rm s}$ is a pure
toroidal component. It is precisely the magnetic pressure at the stellar surface
($\sim B_s^2$) which makes the atmosphere to expand, resulting in a 
magnetic-driven stellar wind. Since radiation pressure is not included here,
the present model illustrates the impact and importance of magnetic effects. 
A more realistic model, combining both, radiation and magnetic pressures and the
contribution of a poloidal component of the field will be the subject of a
future work.

To compute the time evolution of the initial conditions described above, we have
performed the simulations using the magnetohydrodynamic code ZEUS-3D (version
3.4), developed by M. L. Norman and the Laboratory for Computational
Astrophysics. This is a finite-difference, fully explicit, Eulerian code
descended from the code described in Stone \& Norman (1992). A method of
characteristics is used to compute magnetic fields, as described in Clarke
(1996), and flux freezing is assumed in all the runs. We have used spherical
polar coordinates ($r,\theta,\Phi$), with reflecting boundary conditions at the
equator and polar axis. Rotational symmetry is assumed with respect to the polar
axis, and our models are effectively two-dimensional. The simulations are
carried out in the meridional ($r$, $\theta$) plane, but three independent
components of the velocity and magnetic field are computed (\ie the simulations
are ``two and a half'' dimensions).  

\section{Results and Discussion} 

\subsection{Post-AGB Wind Models}

We have first verified our method by using as input stellar conditions the same
parameters used by Pascoli (1997) for an AGB star, namely: $M = 1.1 \Mo$, $R_s =
2 $ A.U., $B_s = 40 $ G, and $\rho_s = 3.5 \times 10^{-10} \gcm $. 

Our grid consists of $200 \times 180$ equidistant zones in $r$ and $\theta$, 
respectively. The innermost radial zone lies at $r_{\rm i}=2 $ A.U., just at the
stellar surface, and the outermost zone at $r_{\rm o} = 80$ A.U. The angular
extent is $90^{\circ}$. For these values, we obtain for the asymptotic velocity
$v_{\infty} = 17.7 \kms$ at $r = 40 R_s$ (i.e., 80 A.U.), in the range 10-20
\kms, in close agreement with the values obtained by Pascoli (1997). Our
solution also agrees in shape with his Figure 3 (case $1/x^2$).

The next step is to select a stellar candidate to compute solutions for post-AGB
winds. We have selected from the literature the well studied object OH 231.8+4.2
(S\'anchez-Contreras et al. 1997; Alcolea et al. 2001; Bujarrabal et al. 2002;
Jura et al. 2002; Desmurs et al.2002), which has a cool (M9III, $T \sim 2000$ K)
central star with $M \approx 1 \Mo$, $R_s = 4.5 $ A.U., and a rotation velocity
of $v_{rot} = 6 \kms $. This object is member of the open cluster NGC 2437
(M46), with an estimated initial mass of $M_{\rm ZAMS} = 3 \Mo $. 

We have computed three wind solutions S1, S2 and S3, with surface magnetic
fields of 0.1, 1 , and 5 G, respectively. The asymptotic terminal velocity has
been measured at $r = 40 R_s$, just above the equatorial plane, obtaining the
following values: 34, 374 and 1,874 $\kms$, respectively. These solutions are
computed with a fixed grid, in which the innermost radial zone lies at 
$r_{\rm i}=4.5$ A.U., just at the stellar surface, and the outermost zone at 
$r_{\rm o}=180$ A.U. The computations stop when the solutions become 
steady-state. These results show the ability of toroidal fields to drive 
stellar winds in the range of a few tens of $\kms$ up to $10^3$ $\kms$.

\subsection{Proto-Planetary Nebula Models}

As a next step we study the type of nebulae that magnetized post-AGB wind models
are able to generate.

A self-expanding grid technique similar to the one used by 
Jun \& Norman (1996) has been used in order to allow growing of the
spatial coordinate by several orders of magnitude. Our expanding grids consist
of $200 \times 180$ equidistant zones in $r$ and $\theta$, respectively. The
initial innermost and outermost radial zones are as described before. These
values are used only up to the point where a shock approaches the outermost
boundary. After that, a shock tracking routine evaluates the expansion velocity
for each forward shock ($v_{\rm s}$) at the polar axis and produces a 
self-expanding grid as $v_g(i)= v_{\rm s} (r(i)/r_{\rm s})$, where $v_g(i)$ and 
$r(i)$ are the velocity and position of each grid zone in the r-coordinate, and
$r_{\rm s}$ is the position of the shock wave. Thus, the final grid size depends
on the dynamical evolution of each individual run. The angular extent is 
$90^{\circ}$ in all cases.

Using this expanding grid technique, we have followed the wind expansion and
nebula formation for six models using the same inputs as before (S1, S2 and S3).
Three of them, models A, B and C have a spherically symmetric initial atmosphere
(Figure 1), while models D, E and F (Figure 2) have an equatorial density
enhancement. To produce this enhancement, we have used the same scheme described
in \cite{GSetal99}, based in the wind compressed zone solutions of Bjorkman \&
Cassinelli (1993), adopting a stellar rotation velocity of $6 \kms$ as observed
in OH 231.8+4.2. The outer radial boundary is updated properly at each loop,
following the expansion of the grid. 

The numerical solutions show that collimation is well established at the very
early phases of evolution, creating jet-like outflows at locations close to
the star. The inclusion of the density enhancement
(Figure 2) produces, as expected, a narrow equatorial waist without any apparent
direct impact at the polar regions. The polar expansion velocities are similar
for all models with the same input magnetic field; models A and D have $v_{exp}
\sim 30  \kms$, models B and E have $v_{exp} \sim 150 \kms$, while models C and
F have $v_{exp} \sim 390 \kms$. 

As a comparative example, Figure 3 compares the result of one of the models 
with 1 G (model E) at 1,000 yr, with two well known, extremely collimated PPNs,
He 3-401 (Sahai 2002) and M 2-9 (Schwarz et al.1997). It is apparent in this
figure that the solution is able to reproduce convincingly the extreme
collimated shapes, along with the sizes and kinematics of these nebulae. 

\subsection{Magnetic Cycles}

Magnetic cycles, and their associated field reversals, have been proposed 
as a plausible origin to the existence of multiple, regularly spaced, and 
faint concentric shells around some planetary nebulae observed with the 
Space Telescope (Soker 2000; Garc\'{\i}a-Segura et al.2001). In fact, OH 
maser observations by
Szymczak et al. (2001) suggest that changes in the polarized maser emission
in some stars could be caused by turbulence in the circumstellar magnetic field
and by global magnetic field reversals. Here, then, we also explore the effects
of magnetic field reversals in magnetic-driven winds, and compare the
results with objects displaying collimated outflows with periodic outburst
features. An interesting example is He 2-90, a PPN whose symmetric and highly
collimated, knotty, bipolar outflow was described by Sahai \& Nyman (2000).
The radial velocities of the knots have been measured by Guerrero et al. (2001),
and the corresponding proper motions subsequently derived by Sahai et al.
(2002). An interesting, and puzzling characteristic in this case is that the
collimated outflow, or jet, maintains a nearly constant apparent width 
throughout
all its extent, i.e. it does not fan out at large distances from the star, and
the velocity of the regularly spaced knots seems to be the same. The ''jet''
speed is somewhere between 150-360 $\kms$, its dynamical time is at least 1400
yr, and the knots are created at the rate of one pair roughly every 35-45 yr.

We have computed model G (similar to E, and shown in Figure 4) with a simple
treatment of the stellar magnetic field ($B_{\rm s}$), allowing it to change
sign in a cycle of the form:
\beq
B_{\rm s}(t) = B_{\rm max} \cos (2 \pi \frac{t}{P}),
\eeq
where $B_{\rm max}$ is the maximum equatorial $B$-field at 
the stellar surface, and $P$ is the period of the magnetic cycle. Since we do
not know the true variation form of the field, this functional form is just a
first order, simple approximation, and is the same scheme used by 
Garc\'{\i}a-Segura et al. (2001) to model the concentric rings around PNs and
PPNs. As in the case of the Sun, we assume that $B_{\rm max}$ has opposite signs
at each hemisphere, with a neutral current sheet near the equatorial plane (\eg
Wilcox \& Ness 1965; Smith, Tsurutani \& Rosenberg 1978). The average thickness
of the neutral current sheet in the Solar case is of about $10^8$ cm, and its
presence does not affect the field outside the equatorial sections. For
simplicity, given that we compute only one hemisphere, we neglect the size of
this current sheet. 
Values for $P$ are adopted arbitrarily in order to match the only source of
observational information available in this regard, namely the estimated 
production of knots.
We have used a period of 80 years for the magnetic cycle
to match the observational data. Thus, every 40 years a new pair of blobs are
produced.

The results of the simulations for model G are shown in Figure 4. It is apparent
from this figure that the similarities of the model with He 2-90 are remarkable.
The highly collimated outflow and its periodic knotty structure are both 
well reproduced, as is the global kinematics (see next section). Thus, in 
spite of the highly
simplified model assumptions, the results indicate that magnetic-driven winds
are able to offer a reasonable explanation to the formation of highly
collimated, knotty bipolar jets with periodic characteristics in PPNs.
Furthermore, the kinematics of the jets that originate from magnetic-driven
winds is different from the one produced by magnetized, line-driven winds.
While magnetized line-driven winds produce observable jets with linearly 
increasing velocity with distance, the so-called Hubble flows (Fig. 5, see also
Garc\'{\i}a-Segura et al. 1999), magnetic-driven winds produce a much more
constant pattern in velocity (Fig. 5). Thus, these differences in the kinematic
behavior could be useful in distinguishing the type of mechanism involved in 
the lunching process of the corresponding outflows.

\subsection{Mechanical energy in the outflows}

We now turn our attention to the kinetic energy and linear momentum contained 
in the outflows from these models. Bujarrabal et al. (2001), as mentioned
earlier, have pointed out that radiation pressure is insufficient to provide the
observed mechanical power in the outflows of PPNs. Figure 6 gives the results
for three different values of the surface magnetic fields covering the initial
1000 years of evolution. The data of PPNs from Bujarrabal et al. (2001) are
indicated as crosses in these plots. The values for most of these objects seen
to be well bracketed by models B (1 G) and C (5 G). Therefore, magnetic-driven
winds are able to provide the necessary energy budget to power the outflows of
PPNs.

\section{Conclusions}  

The origin and evolution of PPNs and PNs represent one of the key questions in
our understanding of stellar physics. Modeling the fascinating features 
displayed by these objects requires not only a better knowledge of stellar
structure at the AGB stage (and beyond) but also a proper consideration of the 
driving mechanisms for mass ejection. The transition from AGB to Post-AGB to PN
central stars involves drastically different conditions at every stage. Whereas
radiation pressure on dust grains is the most likely mechanism at the AGB phase,
as are line-driven winds in the case of PN central stars, for Post-AGB stars the
details of the driving force has been relatively unexplored. A promising avenue,
using dynamo amplification at these late evolutionary stages, has been discussed
by Matt, Frank \& Blackman (2004). They use a simplified model in which the
interface between the (rotating and magnetized) stellar core and envelope stores
a large amount of magnetic energy due to the twisting of an originally poloidal
magnetic field. The magnetic energy is extracted from the stellar rotational 
energy, causing a rapid spin-down of the proto white dwarf, and is able to 
drive a strong and short outburst (this is somehow similar to the "magnetic
bubble" mechanism proposed by Draine 1983, to generate molecular outflows in
star-forming clouds). The outflow can expel the envelope and is termed 
"magnetic explosion" by Matt et al.
 
Indeed, here we do not explore a self-consistent mechanism to generate magnetic
energy but show that a sudden increase of the magnetic field at the onset of the
Post-AGB stage can lead to prominent magnetic-driven stellar winds. These
winds are able to reproduce some of the important characteristics observed in
these transition objects, such as: high mass-loss rates, short transition times
from the late AGB to the PN stage, extreme collimation of the developing nebular
shell and high outflow velocities. In addition, flat kinematic trends in the
outflows and periodic features, as observed in some cases, can also be
explained. Furthermore, the apparently puzzling energy deficit to power the
outflows in this stage is solved if magnetic drivers are considered. Here we
have concentrated in exploring the effects of winds driven by toroidal fields
but future studies will include the contribution of additional ingredients, 
such as radiation pressure and the poloidal magnetic component. Clearly,
more work is required to understand the details of the growth of surface 
magnetic fields at the end of the AGB stage, and their impact in the nebular
structure and morphologies.

{\bf Acknowledgments}

G.G.-S. thanks Noam Soker for fruitful discussions. As usual, we also thank 
Michael L. Norman and the Laboratory for Computational Astrophysics for the use
of ZEUS-3D. The computations were performed at Instituto de 
Astronom\'{\i}a-UNAM. This work has been partially supported by grants from
DGAPA-UNAM (IN130698, IN117799 \& IN114199) and CONACyT (32214-E).

\clearpage

\begin{figure}
\caption{Logarithm of density for models A (0.1 G, left), B (1 G, middle) and 
C (5 G, right) at three different epocs, 50 yr (top), 500 yr (middle) 
and 1000 yr (bottom). Note the change in the spatial scale.
\label{fig1}}
\end{figure}

\clearpage

\begin{figure}
\caption{Logarithm of density for models D (0.1 G, left), E (1 G, middle) 
and F (5 G, right) at three different epocs, 50 yr (top), 500 yr (middle) 
and 1000 yr (bottom). Note the change in the spatial scale. These
models include a equatorial density enhancement.
\label{fig2}}
\end{figure}

\clearpage

\begin{figure}
\caption{Logarithm of density for model E (1 G) at 1000 yr of its
  evolution (center) in comparison with  He 3-401 (Sahai 2002)(left)
  and M 2-9 (Schwarz et al.1997) (right).
\label{fig3}}
\end{figure}

\clearpage

\begin{figure}
\caption{Logarithm of density for model G, a similar model to E , but  
with a magnetic cycle of 80 years.  Each 40 years a pair of blobs are formed at the 
polar axis.
\label{fig4}}
\end{figure}

\clearpage

\begin{figure}
\caption{Logarithm of density (top) and polar velocity (bottom) of three 
different cases. On the left, an example of a magnetized line-driven
wind with sigma = 0.1 (model U in Garc\'{\i}a-Segura et al. 1999).
Middle panels correspond to model E, a magnetic-driven wind. Right panels
correspond to model G, a periodic, magnetic-driven wind.
\label{fig5}}
\end{figure}

\clearpage

\begin{figure}
\caption{Evolution of total linear momentum (top) and total kinetic energy 
(bottom) gained by models A (0.1 G), B (1 G) and C (5 G). Available 
observations (crosses) are taking from Bujarrabal et al.(2001).
\label{fig6}}
\end{figure}

\end{document}